\def\comp{{\rm C}\llap{\vrule height7.1pt width1pt depth-.4pt\phantom t}}
\def\Box{\kern1pt\vbox{\hrule height 1.2pt\hbox{\vrule width 1.2pt\hskip 3pt
   \vbox{\vskip 6pt}\hskip 3pt\vrule width 0.6pt}\hrule height 0.6pt}\kern1pt}
\def\gtwid{\mathrel{\raise.3ex\hbox{$>$\kern-.75em\lower1ex\hbox{$\sim$}}}}
\def\ltwid{\mathrel{\raise.3ex\hbox{$<$\kern-.75em\lower1ex\hbox{$\sim$}}}}
\def\Box{\kern1pt\vbox{\hrule height 1.2pt\hbox{\vrule width 1.2pt\hskip 3pt
   \vbox{\vskip 6pt}\hskip 3pt\vrule width 0.6pt}\hrule height 0.6pt}\kern1pt}
\documentstyle[epsfig,12pt]{article}

\begin{document}
\begin{titlepage}
\begin{flushright}
hep-ph/0007167 \\ CRETE-00-12 \\ UFIFT-HEP-00-12
\end{flushright}
\begin{center}
\textbf{Schwinger's Propagator Is Only A Green's Function}
\end{center}
\begin{center}
N. C. Tsamis$^{\dagger}$
\end{center}
\begin{center}
\textit{Department of Physics \\ University of Crete \\ 710 03 Heraklion, 
Crete \\ GREECE}
\end{center}
\begin{center}
R. P. Woodard$^*$
\end{center}
\begin{center}
\textit{Department of Physics \\ University of Florida \\ 
Gainesville, FL 32611 \\ UNITED STATES}
\end{center}
\begin{center}
ABSTRACT
\end{center}
\hspace*{.5cm} Schwinger used an analytic continuation of the effective action
to correctly compute the particle production rate per unit volume for QED in a 
uniform electric field. However, if one simply evaluates the one loop 
expectation value of the current operator using his propagator, the result is 
zero! We analyze this curious fact from the context of a canonical formalism of
operators and states. The explanation turns out to be that Schwinger's 
propagator is not actually the expectation value of the time-ordered product of
field operators in the presence of a time-independent state, although it is of 
course a Green's function. We compute the true propagator in the presence of a 
state which is empty at $x_+ = 0$ where $x_+ \equiv (x^0+x^3)/\sqrt{2}$ is the 
lightcone evolution parameter. Our result can be generalized to electric fields
which depend arbitrarily on $x_+$.
\begin{flushleft}
\end{flushleft}
\begin{flushleft}
$^{\dagger}$ e-mail: tsamis@physics.uoc.gr \\
$^*$ e-mail: woodard@phys.ufl.edu
\end{flushleft}
\end{titlepage}

\section{Introduction}

Nearly everyone who has made a serious study of quantum field theory in the
past half century is familiar with Schwinger's treatment of quantum 
electrodynamics (QED) in the presence of a constant electromagnetic field 
\cite{Schwinger}. A complicating feature of this background is that particle
production occurs when the electric field is nonzero, so no state can be 
stable even in the free theory. Schwinger was well aware of this feature and
took pains to work around it by assuming, where ever necessary, that the 
electromagnetic field was purely magnetic. He first computed the electron 
propagator then used it to evaluate what would later be known as the in-out
effective action at one loop. He computed the rate of particle production for
the unstable case of an electric field by inferring the imaginary part of the
effective action through an analytic continuation from the stable case of a
magnetic field. 

Schwinger's analysis is rightly regarded as one of the great achievements of
quantum field theory. However, its generality gives rise to a question. 
Schwinger actually obtained a form for the electron propagator for {\it any} 
constant electromagnetic field, magnetic or electric. What would result from 
using his electron propagator to calculate the one loop expectation value of 
the current operator for the unstable case of a constant electric field? This 
is how one might begin computing the back-reaction of particle production on 
the electric field.

The result turns out to be zero. Of course this does not mean that there is no
particle creation at one loop! Rather it implies that, for the case of a 
non-zero electric field, Schwinger's ``propagator'' is really only a Green's 
function and not the expectation value of the time-ordered product of $\psi(x) 
\overline{\psi}(x')$ in the presence of any fixed state. Schwinger never said 
otherwise, and it was presumably to avoid this problem that he employed the 
circuitous analytic continuation procedure.

The distinction between propagators and Green's functions can be understood 
most clearly in the context of the one dimensional harmonic oscillator with 
mass $m$ and frequency $\omega$. We can use the Heisenberg operator equations 
to express the position operator $q(t)$ in terms of its initial values $q_0$ 
and $\dot{q}_0 = p_0/m$,
\begin{equation}
q(t) = q_0 \cos(\omega t) + {p_0 \over m \omega} \sin(\omega t) \; .
\end{equation}
The time-ordered product of two such operators can be expressed in terms of the
$\comp$-number commutator function and the operator anti-commutator,
\begin{eqnarray}
T\{q(t) q(t')\} & = & \frac12 {\rm sgn}({\Delta t}) [q(t),q(t')] + \frac12 
\{q(t),q(t')\} \; , \\
& = & - {i \over 2 m \omega} \sin(m \vert {\Delta t} \vert) + \frac12 \{q(t),
q(t')\} \; , 
\end{eqnarray}
where ${\Delta t} \equiv t - t'$. A general propagator is the expectation value
of this in the presence of some normalized state,
\begin{eqnarray}
i {\Delta}(t;t') & \equiv & \langle S \vert T\{q(t) q(t')\} \vert S \rangle 
\; , \\
& = & -{i \over 2 m \omega} \sin(\omega \vert {\Delta t} \vert) + \alpha
\cos(\omega t) \cos(\omega t') \nonumber \\
& & \qquad + \beta \sin[\omega (t + t')] + \gamma \sin(\omega t) \sin(\omega 
t') \; .
\end{eqnarray}
We have expressed the result in terms of three real numbers $\alpha$, $\beta$ 
and $\gamma$ defined as follows:
\begin{equation}
\alpha \equiv \langle S \vert q_0^2 \vert S \rangle \; , \; 
\beta \equiv {1 \over 2 m \omega} \langle S \vert q_0 p_0 + p_0 q_0 \vert S 
\rangle \; , \; 
\gamma \equiv {1 \over m^2 \omega^2} \langle S \vert p_0^2 \vert S \rangle \; . 
\end{equation}
The key point is that the uncertainty principle imposes an inequality on 
$\alpha$ and $\gamma$,
\begin{equation}
\alpha \gamma \geq \left({1 \over 2 m \omega}\right)^2 \; .
\end{equation}
But $\alpha$, $\beta$ and $\gamma$ can be any $\comp$-numbers if one only 
requires the Green's function equation,
\begin{equation}
-m \left({d^2 \over dt^2} + \omega^2\right) i \Delta(t;t') = i \delta(t - t')
\; ,
\end{equation}
and symmetry under interchange of $t$ and $t'$. For example, $\alpha = \beta = 
\gamma = 0$ gives a Green's function but not a propagator.

This paper contains six sections of which this introduction is the first. In
Section 2 we show that Schwinger's propagator gives zero for the expectation 
value of the current operator at one loop. To see that Schwinger's 
``propagator'' is really only a Green's function it is useful to first express
it in a diagonal function and spinor basis. This is the work of Section 3. In
Section 4 we exhibit an initial value solution for the Heisenberg field
operators analogous to the one presented above for the harmonic oscillator. We 
then show that there is no fixed state which gives Schwinger's result. The 
manner in which it fails also explains the zero current result, and 
incidentally provides what is probably the simplest picture we shall ever get 
of particle production. In Section 5 we work out a true propagator in the 
presence of a natural state. It is significant that we can actually do this for
a class of backgrounds which is general enough to include the actual electric 
field as it evolves under the influence of quantum electrodynamic 
back-reaction. Our conclusions comprise Section 6.

\section{Zero current with Schwinger's propagator}

We begin by representing the propagator as Schwinger did, as the expectation
value of a first quantized operator,
\begin{eqnarray}
i S(x;x') & \equiv & \left\langle x \left\vert {i\over {\rlap{P} /} - e 
{\rlap{A} \thinspace /}(X) - m + i \epsilon} \right\vert x' \right\rangle \; , 
\\
& = & \left\langle x \left\vert \left[{\rlap{P} /} - e {\rlap{A} \thinspace /}
+ m\right] \int_0^{\infty} ds e^{i s [(P - e A)^2 - \frac12 e F_{\mu\nu} 
\sigma^{\mu\nu} - m^2 + i \epsilon]} \right\vert x' \right\rangle \; .
\end{eqnarray}
The position and momentum operators of the first quantized theory are $X^{\mu}$
and $P^{\nu}$, respectively. We assign the standard meaning to $\sigma^{\mu\nu}
\equiv \frac{i}2 [\gamma^{\mu},\gamma^{\nu}]$, and we will assume that the 
vector potential $A_{\mu}(X)$ is a liner function of $X^{\nu}$. We also follow
Schwinger in using the proper time method to regulate expressions involving the
propagator,
\begin{eqnarray}
\lefteqn{iS(x;x') = \lim_{s_0 \rightarrow 0^+} \int_{s_0}^{\infty} ds e^{-i s
(m^2 - i \epsilon)}} \nonumber \\
& & \times \left\langle x \left\vert \left[{\rlap{P} /}-e{\rlap{A}\thinspace /}
+ m\right] e^{- \frac{i}2 e F_{\mu\nu} \sigma^{\mu\nu}} e^{i s [(P - e A)^2} 
\right\vert x' \right\rangle \; .
\end{eqnarray}
Note that if the electric field has magnitude $E$ then the exponential of the
matrix term is,
\begin{equation}
e^{-\frac{i}2 e F_{\mu\nu} \sigma^{\mu\nu}} = \cosh(e E s) I + {1 \over 2 E}
\sinh(e E s) F_{\mu\nu} \gamma^{\mu} \gamma^{\nu} \; .
\end{equation}

At one loop order the expectation value of the current operator is just the
trace of $e \gamma^{\mu}$ up against the coincident propagator,
\begin{eqnarray}
\lefteqn{\left\langle \Omega \left\vert J^{\mu}(x) \right\vert \Omega \right
\rangle = {\rm Tr}\left[e \gamma^{\mu} i S(x;x) \right] \; ,} \\
& = & \lim_{s_0 \rightarrow 0^+} 4 e \int_{s_0}^{\infty} ds e^{-i s (m^2 - i
\epsilon)} \left\{\cosh(e E s) \left\langle x \left\vert [P^{\mu} - e A^{\mu}]
e^{i s (P - e A)^2} \right\vert x \right\rangle \right. \nonumber \\
& & \qquad \qquad \left. - {1 \over E} \sinh(e E s) F^{\mu\nu} \left\langle x
\left\vert [P_{\nu} - e A_{\nu}] e^{i s (P - e A)^2} \right\vert x 
\right\rangle \right\} \; .
\end{eqnarray}
Note that the coincidence limit is regulated by the factor of $e^{is (P-eA)^2}$
as long as $s \neq 0$. Note also that we can get the factor of $P^{\mu} - e A^{
\mu}$ by commutation,
\begin{equation}
P^{\mu} - e A^{\mu} = -\frac{i}2 \left[X^{\mu},(P - e A)^2\right] \; .
\end{equation}
The commutator of $X^{\mu}$ with the exponential gives $is$ times a special 
operator ordering of the product of the exponential times with this factor.
Because the first quantized bra and ket are the same the original ordering can
be restored,
\begin{equation}
\left\langle x \left\vert \left[X^{\mu},e^{i s (P - e A)^2}\right] \right\vert
x \right\rangle = 2 s \left\langle x \left\vert (P^{\mu} - e A^{\mu}) e^{i s 
(P - e A)^2} \right\vert x \right\rangle \; .
\end{equation}
But the commutator vanishes for the same reason. So the expectation value of
the current operator computed using Schwinger's propagator vanishes.

\section{Going to lightcone momentum space}

For definiteness we assume the electric field has magnitude $E$ and is directed
along the positive $z$ axis. If we define the separation 4-vector as ${\Delta x
}^{\mu} \equiv x^{\mu} - x^{\prime \mu}$ then Schwinger's result for the 
electron propagator is \cite{Schwinger}
\begin{eqnarray}
\lefteqn{i S(x;x') = {i \over 32 \pi^2} \exp\left[-i e \int_{x'}^x d\xi^{\mu} 
A_{\mu}(\xi)\right] \int_0^{\infty} ds e^{-i s (m^2 - i \epsilon)}} \nonumber\\
& & \times \exp\left[{i \over 4 s} \left({\Delta x}^2 + {\Delta y}^2 + e E s 
\; {\rm coth}(e E s) [{\Delta z}^2 - {\Delta t}^2]\right) \right] \nonumber \\
& & \qquad \times \left\{\left[-{2 e E \over s} m + { e E \over s^2} \left(
\gamma^1 {\Delta x} + \gamma^2 {\Delta y}\right)\right] \left[{\rm coth}(e E s)
+ \gamma^0 \gamma^3\right] \right. \nonumber \\
& & \qquad \qquad \left. + {e^2 E^2 \over s} {\rm csch}^2(e E s) \left[\gamma^3 
{\Delta z} - \gamma^0 {\Delta t}\right]\right\} \; .
\end{eqnarray}
The special role assumed by the $0$ and $3$ directions strongly suggests that
$i S(x;x')$ should be expressed in terms of lightcone coordinates. The fact 
that translation invariance is broken only by the exponential of the line 
integral of the vector potential also suggests that the propagator should be
transformed to momentum space.

We define the lightcone coordinates and gamma matrices as follows:
\begin{equation}
x_{\pm} \equiv {1 \over \sqrt{2}} \left(x^0 \pm x^3\right) \qquad , \qquad 
\gamma_{\pm} \equiv {1 \over \sqrt{2}} \left(\gamma^0 \pm \gamma^3\right) \; .
\end{equation}
The other (``transverse'') components of $x^{\mu}$ and $\gamma^{\mu}$ comprise
the 2-vectors $\widetilde{x}$ and $\widetilde{\gamma}$, and the invariant 
contraction decomposes as follows,
\begin{equation}
\gamma^{\mu} x_{\mu} = \gamma^0 x^0 - \gamma^3 x^3 - \widetilde{\gamma} \cdot
\widetilde{x} = \gamma_+ x_- + \gamma_- x_+ - \widetilde{\gamma} \cdot 
\widetilde{x} \; .
\end{equation}
Note that $(\gamma_{\pm})^2 = 0$. We follow Kogut and Soper \cite{Kogut} in
defining lightcone spinor projection operators,
\begin{equation}
P_{\pm} \equiv \frac12 \left(I \pm \gamma^0 \gamma^3\right) = \frac12 \gamma_{
\mp} \gamma_{\pm} \; .
\end{equation}
With these conventions the propagator takes the form,
\begin{eqnarray}
i S(x;x') & = & - {i e^2 E^2 \over 32 \pi^2} \exp\left[-i e \int_{x'}^x d\xi^{
\mu} A_{\mu}(\xi)\right] \int_0^{\infty} {ds \over s} {\rm csch}^2(e E s) e^{-i
s (m^2 - i \epsilon)} \nonumber \\
& & \times \exp\left[{i \over 4 s} {\Delta \widetilde{x}} \cdot {\Delta 
\widetilde{x}} - \frac{i}2 e E \; {\rm coth}(e E s) {\Delta x}_+ {\Delta x}_-
\right] \nonumber \\
& & \qquad \times \left\{ \left[{m \over e E} - {\widetilde{\gamma} \cdot 
{\Delta \widetilde{x}} \over 2 e E s}\right] P_+ \left(e^{2 e E s} - 1\right) + 
\gamma_+ {\Delta x}_- \right. \nonumber \\
& & \qquad \qquad \left. + \gamma_- {\Delta x}_+ + \left[{m \over e E} - 
{\widetilde{\gamma} \cdot {\Delta \widetilde{x}} \over 2 e E s}\right] P_- 
\left(1 - e^{-2 e E s}\right) \right\} \; .
\end{eqnarray}

We define the lightcone components of the vector potential $A_{\mu}$ as
\begin{equation}
A_{\pm} \equiv {1 \over \sqrt{2}} \left(A_0 \pm A_3\right) \; .
\end{equation}
Our gauge condition is $A_+ = 0$, and to get $F^{30} = E$ we take $A_- = - E 
x_+$ with $\widetilde{A} = 0$. It is now possible to compute the initial phase.
The path is $\xi^{\mu}(\tau) = x^{\prime\mu} + {\Delta x}^{\mu} \tau$ so we get
\begin{eqnarray}
-i e \int_{x'}^x d\xi^{\mu} A_{\mu}(\xi) & = & i e E {\Delta x}_- \int_0^1 
d\tau \left(x_+' + {\Delta x}_+ \tau\right) \; , \\
& = & \frac{i}2 e E {\Delta x}_- \left(x_+ + x_+'\right) \; .
\end{eqnarray}

The propagator is invariant under translations of $x_-$ and $\widetilde{x}$ so
those are the variables on which we shall Fourier transform,
\begin{eqnarray}
& & i {\widetilde S}\left(x_+,x_+';k_+\widetilde{k}\right) \equiv \int_{-
\infty}^{\infty} d{\Delta x}_- e^{i k_+ {\Delta x}_-} \int d^2{\Delta 
\widetilde{x}} \; e^{-i \widetilde{k} \cdot {\Delta \widetilde{x}}} i S(x;x')
\; , \\
& & = {e^2 E^2 \over 4} \int_0^{\infty} ds \; {\rm csch}^2(e E s) e^{-i s (
\widetilde{\omega}^2 - i \epsilon)} \left\{\left[{m - \widetilde{\gamma} \cdot
\widetilde{k} \over e E}\right] P_+ \left(e^{2 e E s} - 1 \right) \right.
\nonumber \\
& & \qquad \left. -i \gamma_+ {\partial \over \partial k_+} + \gamma_- {\Delta 
x}_+ + \left[{m - \widetilde{\gamma} \cdot \widetilde{k} \over e E}\right] P_- 
\left(1 - e^{-2 e E s}\right)\right\} \nonumber \\
& & \qquad \qquad \times \delta\left(k_+ + \frac12 e E (x_+ + x_+') - \frac12
e E \; {\rm coth}(e E s) {\Delta x}_+\right) \; ,
\end{eqnarray}
where $\widetilde{\omega}^2 = m^2 + \widetilde{k} \cdot \widetilde{k}$. The
delta function can be recast to determine $s$,
\begin{equation}
\delta\left(k_+ + \frac12 e E (x_+ + x_+') - \frac12 e E {\rm coth}(e E s)
{\Delta x}_+\right) = {\delta\left(s - {1 \over 2 e E} \ln\left[{k_+ + e E x_+
\over k_+ + e E x_+'}\right]\right) \over \frac12 e^2 E^2 {\rm csch}^2(e E s)
\vert {\Delta x}_+ \vert} \; .
\end{equation}
This brings the propagator to the form,
\begin{eqnarray}
& & i \widetilde{S}\left(x_+,x_+';k_+\widetilde{k}\right) = {1 \over 2 \vert
{\Delta x}_+ \vert} \left\{\left[{m - \widetilde{\gamma} \cdot \widetilde{k}
\over e E}\right] \left({e E {\Delta x}_+ \over k_+ + e E x_+'}\right) P_+ 
\right. \nonumber \\
& & \qquad \left. - \gamma_+ {\partial \over \partial k_+} + \gamma_- {\Delta 
x}_+ + \left[{m - \widetilde{\gamma} \cdot \widetilde{k} \over e E}\right] 
\left({e E {\Delta x}_+ \over k_+ + e E x_+}\right) P_-\right\} \nonumber \\
& & \qquad \qquad \times \int_0^{\infty} ds e^{-i (\widetilde{\omega}^2 - i
\epsilon)} \delta\left(s - {1 \over 2 e E} \ln\left[{k_+ + e E x_+ \over k_+ +
e E x_+'}\right]\right) \; .
\end{eqnarray}

At this stage it becomes crucial to recall that the electron charge is {\it 
negative} so $e E < 0$. The delta function can only become singular for $s > 0$
if $k_+ > - e E x_+$ for ${\Delta x}_+ > 0$, or if $k_+ > - e E x_+$ for 
${\Delta x}_+ < 0$. The integration over $s$ therefore gives,
\begin{equation}
\left[{k_+ + e E x_+ \over k_+ + e E x_+'}\right]^{{i(\widetilde{\omega}^2 - i
\epsilon) \over -2 e E}} \left\{\theta({\Delta x}_+) \theta\left(k_+ + e E x_+
\right) + \theta(-{\Delta x}_+) \theta\left(-k_+ - e E x_+\right)\right\} \; .
\end{equation}
Note that the $i\epsilon$ in the exponent means that the term in square 
brackets is raised to a power with a small positive real part. This has the 
important consequence that multiplying by $\delta(k_+ + e E x_+)$ gives zero,
so we need not worry about the $-i{\partial}/{\partial k_+}$ acting on the
theta functions. Acting on the power it gives,
\begin{equation}
-i {\partial \over \partial k_+} \left[{k_+ + e E x_+ \over k_+ + e E x_+'}
\right]^{{- i\widetilde{\omega}^2 \over 2 e E}} = {\frac12 \widetilde{\omega}^2
{\Delta x}_+ \over (k_+ +e E x_+) (k_+ + e E x_+')} \left[{k_+ + e E x_+ \over 
k_+ + e E x_+'}\right]^{{- i \widetilde{\omega}^2 \over 2 e E}} \; .
\end{equation}

The final result for the propagator is,
\begin{eqnarray}
\lefteqn{i \widetilde{S}\left(x_+,x_+';k_+,\widetilde{k}\right) =
{\rm sgn}({\Delta x}_+) \; \theta\left[{\rm sgn}({\Delta x}_+) (k_+ + e E x_+)
\right]} \nonumber \\
& & \times \left[{k_+ + e E x_+ \over k_+ + e E x_+'}\right]^{-{i \widetilde{
\omega}^2 \over 2 e E}} \frac12 \left\{\left({m - \widetilde{\gamma} \cdot
\widetilde{k} \over k_+ + e E x_+'}\right) P_+ \right. \nonumber \\
& & \qquad \left. + {\frac12 \widetilde{\omega}^2 \gamma_+ \over(k_+ + e E x_+)
(k_+ + e E x_+')} + \gamma_- + \left({m - \widetilde{\gamma} \cdot\widetilde{k}
\over k_+ + e E x_+}\right) P_- \right\} \; .
\end{eqnarray}
It is more illuminating for the work of the next section to right-multiply this
by $\gamma^0 = (\gamma_+ + \gamma_-)/\sqrt{2}$ and slightly re-arrange the 
order in which the four spinor matrices appear,
\begin{eqnarray}
\lefteqn{i \widetilde{S}\left(x_+,x_+';k_+,\widetilde{k}\right) \gamma^0 = 
{\rm sgn}({\Delta x}_+) \; \theta\left[{\rm sgn}({\Delta x}_+) (k_+ + e E x_+)
\right]} \nonumber \\
& & \times \left[{k_+ + e E x_+ \over k_+ + e E x_+'}\right]^{-{i \widetilde{
\omega}^2 \over 2 e E}} \left\{{1 \over \sqrt{2}} P_+ + {1 \over \sqrt{2}} P_+ 
\frac12 \gamma_- \left({m + \widetilde{\gamma} \cdot \widetilde{k} \over k_+ +
e E x_+'} \right) \right. \nonumber \\
& & \qquad \left. + \left({m - \widetilde{\gamma} \cdot \widetilde{k} \over k_+
+ e E x_+}\right) \frac12 \gamma_+ {1 \over \sqrt{2}} P_+ + {\frac12 
\widetilde{\omega}^2 {1 \over \sqrt{2}} P_- \over (k_+ + e E x_+) (k_+ + e E 
x_+')} \right\} \; . \qquad \label{eq:finalS}
\end{eqnarray}

\section{Inferring the state}

The point of this section is to understand Schwinger's propagator in terms of
operators and states. Let us start with notation for the Fourier transform (on 
$x_-$ and $\widetilde{x}$) of the electron field operator $\psi\left(x_+,x_-,
\widetilde{x}\right)$,
\begin{equation}
\Psi\left(x_+,k_+,\widetilde{k}\right) \equiv \int dx_- e^{i k_+ x_-} \int 
d^2\widetilde{x} e^{-i \widetilde{k} \cdot \widetilde{x}} \psi\left(x_+,x_-,
\widetilde{x}\right) \; .
\end{equation}
This section is about finding a state $\vert S \rangle$ such that,
\begin{eqnarray}
\lefteqn{ i \widetilde{S}\left(x_+,x_+';k_+,\widetilde{k}\right) \gamma^0 
(2\pi)^3 \delta(k_+ -q_+) \delta^2(\widetilde{k} - \widetilde{q})} \nonumber \\
& & = \theta({\Delta x}_+) \left\langle S \left\vert \Psi\left(x_+,k_+,
\widetilde{k}\right) \Psi^{\dagger}\left(x_+',q_+,\widetilde{q}\right) 
\right\vert S \right\rangle \nonumber \\
& & \qquad \qquad - \theta(- {\Delta x}_+) \left\langle S \left\vert \Psi^{
\dagger}\left(x_+',q_+,\widetilde{q}\right) \Psi\left(x_+,k_+,\widetilde{k}
\right) \right\vert S \right\rangle \; , \\
& & \equiv \left\langle S \left\vert X_+\left\{\Psi\left(x_+,k_+,\widetilde{k}
\right) \Psi^{\dagger}\left(x_+',q_+,\widetilde{q}\right)\right\} \right\vert S
\right\rangle \; .
\end{eqnarray}
Note that the adjoint is taken {\it after} Fourier transforming.

We define the $\pm$ components of the electron field (and its Fourier 
transform) using the $P_{\pm}$ projectors,
\begin{equation}
\psi_{\pm}\left(x_+,x_-,\widetilde{x}\right) \equiv P_{\pm} \psi\left(x_+,x_-,
\widetilde{x}\right) \; .
\end{equation}
By acting $P_+$ on the left and right of expression (\ref{eq:finalS}) it is
easy to see that the state $\vert S \rangle$ must obey
\begin{eqnarray}
\lefteqn{\left\langle S \left\vert X_+\left\{\Psi_+\left(x_+,k_+,\widetilde{k}
\right) \Psi_+^{\dagger}\left(x_+',q_+,\widetilde{q}\right)\right\} \right\vert
S \right\rangle = (2\pi)^3 \delta(k_+ - q_+) \delta^2(\widetilde{k} -
\widetilde{q})} \nonumber \\
& & \times {\rm sgn}({\Delta x}_+) \theta\left[{\rm sgn}({\Delta x}_+) (k_+ + 
e E x_+)\right] \left[{k_+ + e E x_+ \over k_+ + e E x_+'} \right]^{- {i 
\widetilde{\omega}^2 \over 2 e E}} \; \times \; {1 \over \sqrt{2}} P_+ \; .
\qquad \label{eq:++}
\end{eqnarray}
The other components can be recognized similarly,
\begin{eqnarray}
\left\langle S \left\vert X_+\left\{\Psi_+ \Psi_-^{\dagger}\right\} \right\vert
S \right\rangle & = & {\rm same} \times {1 \over \sqrt{2}} P_+ \frac12 \gamma_-
\left({m + \widetilde{\gamma} \cdot \widetilde{k} \over k_+ + e E x_+'}\right)
\; , \label{eq:+-} \\
\left\langle S \left\vert X_+\left\{\Psi_- \Psi_+^{\dagger}\right\} \right\vert
S \right\rangle & = & {\rm same} \times \left({m - \widetilde{\gamma} \cdot 
\widetilde{k} \over k_+ + e E x_+'}\right) \frac12 \gamma_+ {1 \over \sqrt{2}}
P_+ \; , \\
\left\langle S \left\vert X_+\left\{\Psi_- \Psi_-^{\dagger}\right\} \right\vert
S \right\rangle & = & {\rm same} \times {\frac12 \widetilde{\omega}^2 {1 \over
\sqrt{2}} P_- \over (k_+ + e E x_+) (k_+ + e E x_+')} \; . \label{eq:--}
\end{eqnarray}

Of relations (\ref{eq:++}-\ref{eq:--}) only the first is really independent,
the others follow from the equations of motion. To see this consider the Dirac
equation for our vector potential,
\begin{equation}
\left(\gamma^{\mu} i\partial_{\mu} - \gamma^{\mu} e A_{\mu} - m\right) 
\psi = \left(\gamma_+ i \partial_+ + \gamma_- (i\partial_- + e E x_+) +
\widetilde{\gamma} \cdot i \widetilde{\nabla} - m \right) \psi \; .
\end{equation}
Multiplication alternately with $\gamma_-$ and $\gamma_+$ gives 
\begin{eqnarray}
i\partial_+ \psi_+(x_+,x_-,\widetilde{x}) & = & \left(m + \widetilde{\gamma} 
\cdot i\widetilde{\nabla} \right) \frac12 \gamma_- \psi_-(x_+,x_-,\widetilde{
x}) \; , \qquad \label{eq:psi+} \\
\left({\mbox{} \over \mbox{}} i\partial_- + e E x_+\right) \psi_-(x_+,x_-,
\widetilde{x}) & = & \left(m + \widetilde{\gamma} \cdot i \widetilde{\nabla} 
\right) \frac12 \gamma_+ \psi_+(x_+,x_-,\widetilde{x}) \; . \qquad 
\label{eq:psi-}
\end{eqnarray}
Fourier transforming (\ref{eq:psi+}) and multiplying by $(m -\widetilde{\gamma}
\cdot \widetilde{k}) \gamma_+/\widetilde{\omega}^2$ gives,
\begin{equation}
\Psi_-\left(x_+,k_+,\widetilde{k}\right) = \left({m - \widetilde{\gamma} \cdot
\widetilde{k} \over \widetilde{\omega}^2}\right) \gamma_+ i \partial_+ \Psi_+
\left(x_+,k_+,\widetilde{k}\right) \; .
\end{equation}
It follows that $\Psi_-$ can be eliminated in favor of $\Psi_+$,
\begin{eqnarray}
\left\langle S \left\vert X_+\left\{\Psi_+ \Psi_-^{\dagger}\right\} \right\vert
S \right\rangle & = & \left\langle S\left\vert X_+\left\{\Psi_+ (-i\partial_+')
\Psi_+^{\dagger}\right\} \right\vert S \right\rangle \gamma_- \left({m + 
\widetilde{\gamma} \cdot \widetilde{q} \over \widetilde{\omega}^2 }\right) \;
, \qquad \label{eq:gen+-} \\
\left\langle S \left\vert X_+\left\{\Psi_- \Psi_+^{\dagger}\right\} \right\vert
S \right\rangle & = & \left({m - \widetilde{\gamma} \cdot \widetilde{k} \over 
\widetilde{\omega}^2}\right) \gamma_+ \left\langle S \left\vert X_+ \left\{i 
\partial_+ \Psi_+ \Psi_+^{\dagger}\right\} \right\vert S \right\rangle \; , \\
\left\langle S \left\vert X_+\left\{\Psi_- \Psi_-^{\dagger}\right\} \right\vert
S \right\rangle & = & \left({m - \widetilde{\gamma} \cdot \widetilde{k} \over 
\widetilde{\omega}^2}\right) \gamma_+ \left\langle S \left\vert X_+ \left\{
\partial_+ \Psi_+ \partial_+' \Psi_+^{\dagger}\right\} \right\vert S 
\right\rangle \nonumber \\
& & \qquad \qquad \qquad \times \gamma_- \left({m + \widetilde{\gamma} \cdot 
\widetilde{q} \over \widetilde{\omega}^2 }\right) \label{eq:gen--} \; .
\end{eqnarray}
The derivatives with respect to $x_+$ and $x_+'$ can be pulled outside the 
$x_+$--ordering symbol if we agree that they do not act on $\theta(\pm {\Delta
x}_+)$. The delta functions in (\ref{eq:++}) set $q_+ = k_+$ and $\widetilde{q}
= \widetilde{k}$ and the derivatives give,
\begin{eqnarray}
i \partial_+ \left[{k_+ + e E x_+ \over k_+ + e E x_+'}\right]^{-{i \widetilde{
\omega}^2 \over 2 e E}} & = & {\frac12 \widetilde{\omega}^2 \over k_+ + e E 
x_+} \left[{k_+ + e E x_+ \over k_+ + e E x_+'}\right]^{-{i \widetilde{\omega
}^2 \over 2 e E}} \; , \\
-i \partial_+' \left[{k_+ + e E x_+ \over k_+ + e E x_+'}\right]^{-{i 
\widetilde{\omega}^2 \over 2 e E}} & = & {\frac12 \widetilde{\omega}^2 \over 
k_+ + e E x_+'} \left[{k_+ + e E x_+ \over k_+ + e E x_+'}\right]^{-{i 
\widetilde{\omega }^2 \over 2 e E}} \; .
\end{eqnarray}
So relations (\ref{eq:+-}-\ref{eq:--}) indeed follow from (\ref{eq:++}).

To identify the state $\vert S \rangle$ we must express the operators as 
functions of $(x_+,k_+,\widetilde{k})$ and functionals of the initial value 
operators. These initial value operators are the only true degrees of freedom
of any Heisenberg operator. On the lightcone the ``initial value surface'' can
be taken as $x_+ = 0$ with $x_- > L$ and $x_+ > 0$ with $x_- = L$. Taking $L$
to $-\infty$ gives the following result for QED in this background 
\cite{Greeks},
\begin{eqnarray}
\lefteqn{\Psi_+\left(x_+,k_+,\widetilde{k}\right) = \left[{k_+ + e E x_+ + i
\epsilon \over k_+ + i \epsilon}\right]^{i \lambda} \Xi_0\left(k_+,
\widetilde{k}\right)} \nonumber \\
& & - \theta(k_+) \theta(-e E x_+ - k_+) \left[{k_+ + e E x_+ + i \epsilon 
\over i \epsilon}\right]^{i \lambda} {\sqrt{2 \pi \lambda} \over \Gamma(1 + i
\lambda)} \Xi_{\infty}\left(k_+,\widetilde{k}\right) \; , \qquad \label{eq:sol}
\end{eqnarray}
where $\lambda \equiv -\widetilde{\omega}^2/(2 e E)$ and the initial value
operators are,
\begin{eqnarray}
\Xi_0\left(k_+,\widetilde{k}\right) & \equiv & \Psi_+\left(0,k_+,\widetilde{k}
\right) \; , \\
\Xi_{\infty}\left(k_+,\widetilde{k}\right) & \equiv & \sqrt{2\pi \over \lambda}
\left({m - \widetilde{\gamma} \cdot \widetilde{k} \over - e E}\right) \frac{i}2
\gamma_- \nonumber \\
& & \times \int d^2\widetilde{x} e^{-i \widetilde{k} \cdot \widetilde{x}} 
\lim_{x_- \rightarrow -\infty} e^{-i k_+ x_-} \psi_-\left(-{k_+ \over e E},
x_-,\widetilde{x}\right) \; .
\end{eqnarray}
The initial value operators anti-commute canonically with their adjoints,
\begin{eqnarray}
\left\{\Xi_0\left(k_+,\widetilde{k}\right),\Xi_0^{\dagger}\left(q_+,\widetilde{
q}\right)\right\} & = & (2\pi)^3 \delta(k_+ - q_+) \delta^2(\widetilde{k} - 
\widetilde{q}) {1 \over \sqrt{2}} P_+ \; , \label{eq:AC0} \\ 
\left\{\Xi_{\infty}\left(k_+,\widetilde{k}\right),\Xi_{\infty}^{\dagger}\left(
q_+,\widetilde{q}\right)\right\} & = & (2\pi)^3 \delta(k_+ - q_+) \delta^2(
\widetilde{k} - \widetilde{q}) {1 \over \sqrt{2}} P_+ \; . \label{eq:ACinf}
\end{eqnarray}
The ``$0$`` operators anti-commute with the ``$\infty$'' ones by causality 
since the two surfaces are spacelike related for all finite $L$.

It is worth pausing at this point to comment on the significant features of our
operator solution. First, note that for $k_+ > 0$ the mode functions experience 
a characteristic drop in amplitude as they evolve through the singular point
at $x_+ = -k_+/eE$,
\begin{equation}
\left\vert \left[k_+ + e E x_+ + i \epsilon \over k_+ + i \epsilon\right]^{i 
\lambda} \right\vert = \cases{1 & $\forall x_+ < -k_+/eE$ \cr e^{-\pi \lambda} 
& $\forall x_+ > - k_+/eE$} \; . \label{eq:amp}
\end{equation}
The amplitude lost by $\Xi_0\left(k_+,\widetilde{k}\right)$ passes to 
$\Xi_{\infty}\left(k_+,\widetilde{k}\right)$ by virtue of the identity \cite{
Russians},
\begin{equation}
\left\vert \left[{k_+ + e E x_+ + i \epsilon \over i \epsilon}\right]^{i 
\lambda} {\sqrt{2\pi \lambda} \over \Gamma(1 + i \lambda)}\right\vert =
\sqrt{1 - e^{-2\pi \lambda}} \; ,
\end{equation}
for $0 < k_+ < - e E x_+$. The physical interpretation of the amplitude drop is
particle production. This is a discrete and instantaneous event on the 
lightcone. As each mode passes through its singularity at $x_+ = -k_+/eE$ the
eigenvalue of $-i\partial_+$ changes sign and the operator coefficient switches
interpretation from annihilator to creator. Since Heisenberg states are fixed 
a state which was originally empty in that mode seems, after singularity, to
have acquired a particle with probability equal to the square of the 
post-singular amplitude (\ref{eq:amp}).

The second point worthy of mention about (\ref{eq:sol}) is that we cannot 
follow the usual lightcone practice, for nonzero mass and/or more than two 
spacetime dimensions, of ignoring the ``$\infty$'' operators. These are always
{\it technically} present in lightcone field theory, but they usually remain 
segregated to sector at $k_+ = 0$. In computing scattering amplitudes one can 
ignore this sector and recover the zero momentum limit instead by analytic 
continuation. We cannot get away with this for QED in a constant electric 
field. The {\it physical} momentum is the minimally coupled one, $p_+ \equiv 
k_+ + e E x_+$, which reaches the far infrared near singularity. At this point 
operators from the surface at $x_- = -\infty$ can and do mix with the ``$0$'' 
operators in (\ref{eq:sol}).

We should end this digression by noting that part of our operator solution 
(\ref{eq:sol}) has been obtained in a different context by Srinivasan and 
Padmanabhan \cite{Indians1,Indians2}. The mode solution they give corresponds 
to the term proportional to $\Xi_0\left(k_+,\widetilde{k}\right)$, although 
without our $i \epsilon$ convention. They do not get the part proportional to 
$\Xi_{\infty} \left(k_+,\widetilde{k}\right)$. They employed a WKB approach to 
evolve around the singularity at $k_+ + e E x_+ = 0$, and they claim this 
results in a second term proportional to $\Xi_0\left(-k_+,\widetilde{k}
\right)$. We have been unable to understand why the WKB technique applies to a 
first order evolution equation, nor can we reproduce their results.

Our operator solution (\ref{eq:sol}) implies the following anti-commutation 
relation for $x_+ \neq x_+'$:
\begin{equation}
\left\{\Psi_+\left(x_+,k_+\widetilde{k}\right),\Psi_+^{\dagger}\left(x_+',q_+,
\widetilde{q}\right)\right\} = (2\pi)^3 \delta^3(k - q) \left\vert {k_+ + e E
x_+ \over k_+ + e E x_+'}\right\vert^{i \lambda} {1 \over \sqrt{2}} P_+ \; .
\end{equation}
Comparison with (\ref{eq:++}) reveals that $\vert S \rangle$ must obey,
\begin{eqnarray}
\left\langle S \left\vert \Psi_+\left(x_+,k_+,\widetilde{k}\right) ,\Psi_+^{
\dagger}\left(x_+',q_+,\widetilde{q}\right) \right\vert S \right\rangle & = & 
\theta(k_+ + e E x_+) \left\{\Psi_+,\Psi_+^{\dagger}\right\} \; , \qquad \\
\left\langle S \left\vert \Psi_+^{\dagger}\left(x_+',q_+,\widetilde{q}\right) ,
\Psi_+\left(x_+,k_+,\widetilde{k}\right) \right\vert S \right\rangle & = & 
\theta(- k_+ - e E x_+) \left\{\Psi_+,\Psi_+^{\dagger}\right\} \; . \qquad
\end{eqnarray}
The state $\vert S \rangle$ must therefore be annihilated by $\Psi_+$ for $k_+
+ e E x_+ > 0$ and by $\Psi_+^{\dagger}$ for $k_+ + e E x_+ < 0$. In terms of
the fundamental operators this translates to,
\begin{eqnarray}
\Xi_0\left(k_+ + e E x_+,\widetilde{k}\right) \left\vert S \right\rangle = & 0 
& = \Xi_0^{\dagger}\left(-k_+ - e E x_+,-\widetilde{k}\right) \left\vert S 
\right\rangle \; , \label{eq:S0} \\
\Xi_{\infty}\left(k_+ + e E x_+,\widetilde{k}\right) \left\vert S \right\rangle
= & 0 & = \Xi_{\infty}^{\dagger}\left(-k_+ - e E x_+,-\widetilde{k}\right) 
\left\vert S \right\rangle \; , \label{eq:Sinf}
\end{eqnarray}
for all $k_+ > 0$. It is immediately apparent that $\vert S \rangle$ is not a
proper Heisenberg state in the sense of remaining fixed. We must rather use a
{\it different} state for each value of $x_+$ in order to recover Schwinger's
result. Hence it is only a Green's function and not a true propagator.

We can also understand the curious result of Section 2 that Schwinger's 
``propagator'' gives a null result for the expectation vlaue of the current 
operator at one loop. Recall from our operator solution that the eigenvalue of
$-i\partial_+$ on $\Psi_+\left(x_+,k_+,\widetilde{k}\right)$ is negative for
$k_+ + e E x_+ > 0$ and positive for $k_+ + e E x_+ < 0$. This means that at 
fixed $x_+$ the electron annihilators are proportional to $\Psi_+\left(x_+,k_+,
\widetilde{k}\right)$, for all $k_+ + e E x_+ > 0$, and the positron 
annihilators are proportional to $\Psi_+^{\dagger}\left(x_+,k_+,\widetilde{k}
\right)$, again for $k_+ + e E x_+ > 0$. Conditions (\ref{eq:S0}-\ref{eq:Sinf})
guarantee that these operators annihilate $\vert S \rangle$, which means that
the state is empty at $x_+$. This is no contradiction with the fact that 
particle production really happens in this background because it is always 
possible to find a state which is empty at one particular instant. If the state
were held fixed one would see a nonzero (actually infinite) current at later 
$x_+$ \cite{Greeks}, but that is not what Schwinger's ``propagator'' does. As 
$x_+$ changes the state also changes to the one which happens to be 
instantaneously empty at the new value of $x_+$. So one of course sees zero 
current. End of mystery.

\section{A true propagator}

We have been able to solve the Heisenberg operator equations for a vector 
potential $A_-(x_+)$ which depends arbitrarily upon $x_+$.\footnote{At certain 
points we do assume that $e A_-(x_+)$ is an {\it increasing} function of $x_+$.
This could be avoided at the cost of more complicated expressions.} Since there
is no significant simplification arising from the assumption that $A_-(x_+) =
- E x_+$, we begin by stating the more general solution. The derivation can be 
found elsewhere \cite{Greeks}.

Modes still undergo particle production when the minimally coupled momentum
$k_+ - e A_-(x_+)$ vanishes. We define this value of $x_+$ as $X(k_+)$,
\begin{equation}
k_+ = A_-\left(X(k_+)\right) \; .
\end{equation}
Since the electric field is no longer necessarily constant we must generalize
the definition of $\lambda$,
\begin{equation}
\lambda\left(k_+,\widetilde{k}\right) \equiv {\widetilde{\omega}^2 \over 2 e 
A_-'(X(k_+))} \; .
\end{equation}
The mode functions require a similar generalization,
\begin{equation}
{\cal E}[A_-]\left(y_+,x_+;k_+,\widetilde{k}\right) \equiv \exp\left[-\frac{i}2
\widetilde{\omega}^2 \int_{y_+}^{x_+} {du \over k_+ - e A_-(u) + i \epsilon}
\right] \; .
\end{equation}
With these conventions the operator solution can be written as 
follows:\footnote{We have actually quoted a simplified form in which a 
distributional limit was taken assuming that $k_+$ is somewhat displaced from 
the singular points at $k_+ = 0$ and $k_+ = e A_-(x_+)$. The more accurate
expression must be used in taking derivatives with respect to $x_+$ 
\cite{Greeks}.}
\begin{eqnarray}
\lefteqn{\Psi_+\left(x_+,k_+,\widetilde{k}\right) = {\cal E}[A_-]\left(0,x_+;
k_+,\widetilde{k}\right) \Xi_0\left(k_+,\widetilde{k}\right)} \nonumber \\
& & - \theta(k_+) \theta(e A_-(x_+) - k_+) {\cal E}\left(X(k_+),x_+;k_+,
\widetilde{k}\right) {\sqrt{2 \pi \lambda} \over \Gamma(1 + i \lambda)} 
\Xi_{\infty}\left(k_+,\widetilde{k}\right) \; , \qquad \label{eq:newsol}
\end{eqnarray}
where the initial value operators are,
\begin{eqnarray}
\Xi_0\left(k_+,\widetilde{k}\right) & \equiv & \Psi_+\left(0,k_+,\widetilde{k}
\right) \; , \\
\Xi_{\infty}\left(k_+,\widetilde{k}\right) & \equiv & \sqrt{2\pi \over \lambda(
k_+,\widetilde{k})} \left({m - \widetilde{\gamma} \cdot \widetilde{k} \over e 
A_-'(X(k_+))}\right) \frac{i}2 \gamma_- \nonumber \\
& & \times \int d^2\widetilde{x} e^{-i \widetilde{k} \cdot \widetilde{x}} 
\lim_{x_- \rightarrow -\infty} e^{-i e k_+ x_-} \psi_-\left({\mbox{} \over
\mbox{}} X(k_+),x_-,\widetilde{x}\right) \; . \qquad
\end{eqnarray}
The canonical anti-commutation relations (\ref{eq:AC0}-\ref{eq:ACinf}) are 
unchanged.

It is natural to study the state that is empty at $x_+ = 0$, which means
\begin{equation}
\Xi_0\left(k_+,\widetilde{k}\right) \vert \Omega \rangle = 0 = 
\Xi_0^{\dagger}\left(-k_+,-\widetilde{k}\right) \vert \Omega \rangle \; ,
\end{equation}
for all $k_+ > 0$. It is also natural to forbid particles from entering via the
surface at $x_- = -\infty$, which implies
\begin{equation}
\Xi_{\infty}^{\dagger}\left(k_+,\widetilde{k}\right) \vert \Omega \rangle = 0
\; ,
\end{equation}
also for $k_+ > 0$. The two operator orderings of $\Psi_+$ and $\Psi_+^{\dagger
}$ give,
\begin{eqnarray}
& & \left\langle \Omega \left\vert \Psi_+\left(x_+,k_+,\widetilde{k}\right) 
\Psi_+^{\dagger}\left(x_+',q_+,\widetilde{q}\right) \right\vert \Omega 
\right\rangle = (2\pi)^3 \delta(k_+  - q_+) \delta^2\left(\widetilde{k} -
\widetilde{q}\right) {P_+ \over \sqrt{2}} \nonumber \\
& & \qquad \qquad \qquad \times \theta(k_+) {\cal E}\left(0,x_+;k_+,\widetilde{
k}\right) {\cal E}^*\left(0,x_+';k_+,\widetilde{k}\right) \; , \\
& & \left\langle \Omega \left\vert \Psi_+^{\dagger}\left(x_+',q_+,\widetilde{q}
\right) \Psi_+\left(x_+,k_+,\widetilde{k}\right) \right\vert \Omega 
\right\rangle = (2\pi)^3 \delta(k_+ - q_+) \delta^2\left(\widetilde{k} - 
\widetilde{q}\right) {P_+ \over \sqrt{2}} \nonumber \\
& & \; \times \left\{{\mbox{} \over \mbox{}} \theta(k_+) \theta\left(x_+ -
X\right) {\cal E}\left(X,x_+;k_+,\widetilde{k}\right) {\cal E}^*\left(X,x_+';
k_+,\widetilde{k}\right) \left(e^{\pi \lambda} - e^{-\pi \lambda}\right)
\right. \nonumber \\
& & \qquad \qquad \qquad \left. {\mbox{} \over \mbox{}} + \theta(-k_+) 
{\cal E}\left(0,x_+;k_+,\widetilde{k} \right) {\cal E}^*\left(0,x_+';k_+,
\widetilde{k}\right) \right\} \; .
\end{eqnarray}

Now note that the conjugated mode function can be re-expressed using the Dirac
identity,
\begin{eqnarray}
\lefteqn{{\cal E}^*\left(0,x_+';k_+,\widetilde{k}\right) = \exp\left[\frac{i}2
\widetilde{\omega}^2 \int_0^{x_+'} {du \over k_+ - e A_-(u) - i \epsilon}
\right] \; , } \\
& & \longrightarrow \exp\left[\frac{i}2 \widetilde{\omega}^2 \int_0^{x_+'} du
\left\{{1 \over k_+ - e A_-(u) + i \epsilon} + 2\pi i \delta\left(k_+ -e A_-(u)
\right)\right\}\right] \; , \qquad \\
& & = \exp\left[\frac{i}2 \widetilde{\omega}^2 \int_0^{x_+'} {du \over k_+ - e
A_-(u) + i \epsilon}\right] e^{-2\pi \lambda \theta(k_+) \theta(x_+'-X)} \; .
\end{eqnarray}
We can therefore write the various ${\cal E} \times {\cal E}^*$ products as,
\begin{eqnarray}
\lefteqn{{\cal E}\left(0,x_+;k_+,\widetilde{k}\right) {\cal E}^*\left(0,x_+';i
k_+,\widetilde{k}\right)} \nonumber \\
& & \qquad \qquad \longrightarrow {\cal E}\left(x_+',x_+;k_+\widetilde{
k}\right) e^{-2 \pi \lambda \theta(k_+) \theta(x_+' - X)} \; , \qquad \\
\lefteqn{{\cal E}\left(X,x_+;k_+,\widetilde{k}\right) {\cal E}^*\left(X,x_+';
k_+,\widetilde{k}\right)} \nonumber \\
& & \qquad \qquad \longrightarrow {\cal E}\left(x_+',x_+;k_+\widetilde{
k}\right) e^{-\pi \lambda \theta(k_+) \theta(x_+' - X)} \; . \qquad
\end{eqnarray}
Assembling all the pieces gives the following result for the expectation value
of the $x_+$--ordered product:
\begin{eqnarray}
\lefteqn{\left\langle \Omega \left\vert X_+\left\{\Psi_+\left(x_+,k_+,
\widetilde{k}\right) \Psi_+^{\dagger}\left(x_+',q_+,\widetilde{q}\right)
\right\} \right\vert \Omega \right\rangle = (2\pi)^3 \delta^3(k - q) 
{P_+ \over \sqrt{2}}} \nonumber \\
& & \times {\cal E}\left(x_+',x_+;k_+\widetilde{k}\right) \left\{\theta({\Delta
x}_+) \theta(k_+) e^{-2 \pi \lambda \theta(x_+' - X)} - \theta(-{\Delta x}_+) 
\theta(-k_+) \right. \nonumber \\
& & \qquad \left. - \theta(-{\Delta x}_+) \theta(k_+) \theta(x_+ - X) \left(1 -
e^{-2 \pi \lambda}\right) \right\} \; .
\end{eqnarray}
Expanding the exponential which contains the theta function,
\begin{equation}
e^{-2 \pi \lambda \theta(x_+' - X)} = \theta(X - x_+') + \theta(x_+' - X) e^{-2
\pi \lambda} \; ,
\end{equation}
and simplifying under the assumption that $e A_-(u)$ is an increasing function
gives the following more compact result,
\begin{eqnarray}
\lefteqn{\left\langle \Omega \left\vert X_+\left\{\Psi_+\left(x_+,k_+,
\widetilde{k}\right) \Psi_+^{\dagger}\left(x_+',q_+,\widetilde{q}\right)
\right\} \right\vert \Omega \right\rangle = (2\pi)^3 \delta^3(k - q) 
{P_+ \over \sqrt{2}}} \nonumber \\
& & \times {\cal E}\left(x_+',x_+;k_+\widetilde{k}\right) \left\{\theta({\Delta
x}_+) \theta(X - x_+') - \theta(-{\Delta x}_+) \theta(X - x_+) \right. 
\nonumber \\
& & \qquad \left. + \theta(x_+ - X) \theta(x_+' - X) \theta(k_+) e^{-2\pi 
\lambda}\right\} \; .
\end{eqnarray}

We can still obtain the minus components from the plus ones,
\begin{equation}
\Psi_-\left(x_+,k_+,\widetilde{k}\right) = \left({m - \widetilde{\gamma} \cdot
\widetilde{k} \over \widetilde{\omega}^2}\right) \gamma_+ i \partial_+ \Psi_+
\left(x_+,k_+,\widetilde{k}\right) \; ,
\end{equation}
so that relations (\ref{eq:gen+-}-\ref{eq:gen--}) still determine the other 
components of the propagator from the $++$ ones. So the full 2-point function
is,
\begin{eqnarray}
\lefteqn{\left\langle \Omega \left\vert X\left\{\Psi\left(x_+,k_+,\widetilde{
k}\right) \Psi^{\dagger}\left(x_+',q_+,\widetilde{q}\right)\right\} \right\vert
\Omega \right\rangle = (2\pi)^3 \delta(k_+ - q_+) \delta^2\left(\widetilde{k}
- \widetilde{q}\right)} \nonumber \\
& & \times {\cal E}\left(x_+',x_+;k_+\widetilde{k}\right) \left\{\theta({\Delta
x}_+) \theta(X - x_+') - \theta(-{\Delta x}_+) \theta(X - x_+) \right. 
\nonumber \\
& & \quad \left. + \theta(x_+ - X) \theta(x_+' - X) \theta(k_+) e^{-2\pi 
\lambda} \right\} \left\{{1 \over \sqrt{2}} P_+ \right. \nonumber \\
& & \qquad + {P_+ \gamma_- \over 2 \sqrt{2}} \left({m + \widetilde{\gamma} 
\cdot \widetilde{k} \over k_+ - e A_-(x_+') - i \epsilon} \right) + \left({m - 
\widetilde{\gamma} \cdot \widetilde{k} \over k_+ - e A_-(x_+) + i \epsilon}
\right) {\gamma_+ P_+ \over 2 \sqrt{2}} \nonumber \\
& & \qquad \qquad \left. + {\frac12 \widetilde{\omega}^2 {1 \over \sqrt{2}} P_-
\over (k_+ - e A_-(x_+)+ i \epsilon) (k_+ - e A_-(x_+') - i \epsilon)} \right\} 
\; . \label{eq:ourprop}
\end{eqnarray}

\section{Discussion}

We have shown that, if one computes the one loop expectation value of the 
current using Schwinger's propagator, the result is zero. The reason for this
emerges when one attempts to write the propagator as the expectation value of
the time-ordered product of $\psi(x) \overline{\psi}(x')$ in the presence of
some state. The only way to do so is with a state which changes as the time
parameter ($x_+$) of the $\psi$ field does. At each different value of $x_+$ 
the appropriate state is the one which happens to be free of particles at that
instant, so of course the current is zero.

Because of the special roles played by time and by the direction of the 
electric field, lightcone coordinates are particularly well suited to this 
problem. At the level of operators and states this implies a profound departure
from the usual picture in which one imagines specifying a state on a surface of
constant $x^0$. For a state specified instead on a surface of constant $x_+$ 
the phenomenon of pair production is a discrete and instantaneous event. For 
each mode of fixed $k_+$ it occurs when the minimally coupled momentum $p_+ = 
k_+ - e A_(x_+)$ vanishes. This can be understood quite simply by representing
the lightcone system in the standard way as the infinite boost limit of a 
conventional system in which the states are defined on surfaces of constant 
time \cite{Kogut}. Particles produced with finite $p^{\prime 3}$ in that frame 
must have $p^3 = -\infty$ in the lightcone frame, and $p^3= -\infty$ implies 
$p_+ = 0^+$ for a particle which is on shell \cite{Greeks}.

A remarkable feature of our treatment is that explicit (and quite simple!) mode
functions can be obtained for a background in which the electric field depends 
arbitrarily upon the lightcone coordinate $x_+$.\footnote{The simplicity of the
lightcone mode functions has been noted previously, for the special case of 
constant electric field, by Srinivasan and Padmanabhan 
\cite{Indians1,Indians2}. Note, however, that we do not agree with some aspects
of their solution.} We have in fact worked out the actual electron propagator
for such a background in the presence of a state which is empty at $x_+ = 0$.
Since neither the photon propagator nor the vertices of QED are affected by the
background, expression (\ref{eq:ourprop}) completes the Feynman rules with 
which one can compute the quantum electrodynamic back-reaction to any order. It
is significant that we can get the propagator for a class of backgrounds 
general enough to include the actual solution as the electric field evolves 
under the impact of the current it generates.

\vskip 1cm
\centerline{\bf Acknowledgements}

We thank D. Boyanovsky, C. B. Thorn and T. N. Tomaras for discussions on 
related subjects. This work was partially supported by DOE contract 
DE-FG02-97ER\-41029, by the Greek General Secretariat of Research and 
Technology grant 97 E$\Lambda$--120, and by the Institute for Fundamental 
Theory.

\end{document}